\documentclass[conference]{IEEEtran}
\IEEEoverridecommandlockouts
\usepackage{cite}
\usepackage{amsmath,amssymb,amsfonts}
\usepackage{graphicx}
\usepackage{textcomp}
\usepackage{xcolor}
\def\BibTeX{{\rm B\kern-.05em{\sc i\kern-.025em b}\kern-.08em
    T\kern-.1667em\lower.7ex\hbox{E}\kern-.125emX}}

\usepackage{array}
\usepackage[caption=false,font=footnotesize]{subfig}
\usepackage{textcomp}
\usepackage{stfloats}
\usepackage{url}
\usepackage{verbatim}

\usepackage{mathtools}
\usepackage{siunitx}
\usepackage{booktabs}

\usepackage[ruled,vlined]{algorithm2e}
\SetKw{Break}{break}  
\SetKw{Continue}{continue}  

\newcommand{\freal}{\mathbb{R}}

\newcommand{\lEbNO}{E_\textnormal{b} / N_0}

\usepackage{dsfont}
\newcommand{\ind}{\mathds{1}}

\usepackage{bm}

\usepackage{amsthm}

\usepackage{enumitem}

\usepackage{acro}

\DeclareAcronym{ORBGRAND}{
    short = ORBGRAND,
    long = Ordered Reliability Bits Guessing Random Additive Noise Decoding,
}

\DeclareAcronym{LLR}{
    short = LLR,
    long = log likelihood ratio,
}

\DeclareAcronym{BIAWGN}{
    short = BI-AWGN,
    long = binary-input additive white Gaussian noise,
}

\DeclareAcronym{GRAND}{
    short = GRAND,
    long = Guessing Random Additive Noise Decoding,
}

\DeclareAcronym{ML}{
    short = ML,
    long = maximum likelihood,
}

\DeclareAcronym{SGRAND}{
    short = SGRAND,
    long = Soft GRAND,
}

\DeclareAcronym{SOGRAND}{
    short = SOGRAND,
    long = Soft-output GRAND,
}

\DeclareAcronym{SyGRAND}{
    short = SyGRAND,
    long = Syndrome Enhanced GRAND,
}

\DeclareAcronym{BLER}{
    short = BLER,
    long = block error rate,
}

\DeclareAcronym{URLLC}{
    short = URLLC,
    long = Ultra-Reliable Low-Latency Communication,
}

\DeclareAcronym{SNR}{
    short = SNR,
    long = signal-to-noise ratio,
}

\DeclareAcronym{GCD}{
    short = GCD,
    long = Guessing Codeword Decoding,
}

\DeclareAcronym{GRANDAB}{
    short = GRANDAB,
    long = GRAND with abandonment,
}

\DeclareAcronym{ORDEPT}{
    short = ORDEPT,
    long = Ordered Reliability Direct Error Pattern Testing
}

\DeclareAcronym{eBCH}{
    short = eBCH,
    long = extended Bose–Chaudhuri–Hocquenghem,
}

\DeclareAcronym{CRC}{
    short = CRC,
    long = cyclic redundancy check,
}

\DeclareAcronym{CAPolar}{
    short = CA-Polar,
    long = CRC-Assisted Polar,
}

\DeclareAcronym{BPSK}{
    short = BPSK,
    long = binary phase-shift keying,
}

\begin{document}

\title{SOGRAND Assisted Guesswork Reduction 

\author{
\IEEEauthorblockN{Lukas Rapp}
\IEEEauthorblockA{\textit{Research Laboratory for Electronics} \\
\textit{Massachusetts Institute of Technology}\\
Cambridge, USA \\
rappl@mit.edu}
\and
\IEEEauthorblockN{Muriel M{\'e}dard}
\IEEEauthorblockA{\textit{Research Laboratory for Electronics} \\
\textit{Massachusetts Institute of Technology}\\
Cambridge, USA \\
medard@mit.edu}
\and
\IEEEauthorblockN{Ken R. Duffy}
\IEEEauthorblockA{\textit{Dept. ECE \& Math} \\
\textit{Northeastern University}\\
Boston, USA \\
k.duffy@northeastern.edu}
}

}

\maketitle

\begin{abstract}
    Proposals have been made to reduce the guesswork of Guessing Random Additive Noise Decoding (GRAND) for binary linear codes by leveraging codebook structure at the expense of degraded block error rate (BLER). We establish one can preserve guesswork reduction while eliminating BLER degradation through dynamic list decoding terminated based on Soft Output GRAND’s error probability estimate. We illustrate the approach with a method inspired by published literature and compare performance with Guessing Codeword Decoding (GCD). We establish that it is possible to provide the same BLER performance as GCD while reducing guesswork by up to a factor of \(32\).
\end{abstract}


\section{Introduction}
\ac{GRAND}~\cite{duffyCapacityAchievingGuessingRandom2019} is a class of algorithms that can decode any moderate redundancy error correction code of any structure, including nonlinear codes, e.g., \cite{cohenAESErrorCorrection2023}. \ac{GRAND} generates noise effect patterns and queries whether what remains when they are removed from hard decision sequences corresponds to a valid codeword. If GRAND's noise effect query order is from most likely to least likely, the first valid codeword is a \ac{ML} decoding~\cite{duffyCapacityAchievingGuessingRandom2019}. For hard detection systems \cite{duffyCapacityAchievingGuessingRandom2019,An22GRANDMO}, precise query orders can be efficiently created in software and circuits for binary symmetric channels and channels with Markov memory, e.g. \cite{abbas2020grand,abbas2021grandmo,Riaz21,Burg24}. 

For soft detection systems it is possible to produce perfect query orders in software, e.g. \cite{solomonSoftMaximumLikelihood2020b}, but they are not suitable for efficient hardware implementation. For practical implementations, approximations such as \ac{ORBGRAND} \cite{duffyOrderedReliabilityBits2022,an2023soft,duffyUsingChannelCorrelation2023a}, for which a number of syntheses \cite{condo2021_highperformance, condo2021fixed,abbas2022orbgrand} and a taped out chip \cite{Riaz23,Riaz25} exist, have been developed that establish practicality. 
The cost of any approximate query order is that, on occasion, an erroneous decoding can be identified before the ML decoding resulting in degraded \ac{BLER} performance. For \ac{ORBGRAND}, for example, that BLER degradation only occurs at higher SNR. 

One way to rectify that increase in BLER is to use a list-decoding approach, where decoding is not terminated when the first codeword is identified but continues until multiple codewords are identified and then the most likely codeword is selected from that list. Such an approach has been proposed for ORBGRAND \cite{abbasListGRANDPracticalWay2023}. While list decoding improves the BLER, it does so at the cost of an increased number of queries required to identify more than one codeword. As the first codeword found is typically correct, subsequent work can be wasteful, resulting in a trade-off between precision and effort. 

Recently, it has been established that any Soft Input (SI) \ac{GRAND} algorithm, regardless of whether its query order is optimal or not, can readily produce accurate blockwise Soft Output (SO) in the form of an estimate that each element of a decoding list, including a list of size $1$, is correct or that the correct decoding has not been found, resulting in \ac{SOGRAND} \cite{galligan2023upgrade,yuanSOGRAND25}. Such blockwise SO can be used to adaptively terminate list decoding when the confidence in correct decoding is sufficiently high, creating a dynamic approach to list sizing~\cite{yuanSOGRAND25}. With guesswork defined to be the number of queries until a decoding is returned or decoding is abandoned, here we establish that the \ac{SOGRAND} principle can be used with query reduction methods that exploit codebook structure to enhance guesswork savings while removing the BLER penalty by dynamically terminating decoding.

For binary linear codes, proposals have been made that leverage codebook structure to inform the guesswork order. In particular, Rowshan and Yuan~\cite{rowshanConstrainedErrorPattern2022a} pointed out that for even-weight binary codes, such as eBCH and CA-Polar codes, guesswork can be reduced by a factor of up to two without any impact to BLER by only generating queries that have the same parity as the hard decision sequence. More generally, they introduced segmented \ac{GRAND}~\cite{rowshanConstrainedErrorPattern2022a,rowshanLowComplexityGRANDSegmentation2023,rowshanSegmentedGRANDComplexity2025a}, which uses independent noise effect generators on non-overlapping code segments, solely generating queries that necessarily fulfill a subset of the parity-check constraints. As the resulting noise effects are only approximately in order from most likely to least likely, the reduction in average guesswork comes at the cost of higher \ac{BLER}, e.g.~\cite{isitpaper}, which could be overcome with additional guesswork and a list-decoding approach.

Hadavian et al.~\cite{hadavianOrderedReliabilityDirect2023} introduced \ac{ORDEPT}, a distinct method that leverages the syndrome of each guess to find valid candidate codewords that are an additional bit flip away from the current guess. Interpreted from a \ac{GRAND} perspective, \ac{ORDEPT} can be considered as a variant with a suboptimal noise effect generator that yields valid codewords early in the guessing order by incorporating codebook structure. The approach also reduces average guesswork, but leads to a degradation in \ac{BLER} performance that can be mitigated with list decoding at the cost of additional guesswork~\cite{hadavianOrderedReliabilityDirect2023}. 

In this paper, we consider the use of \ac{SOGRAND} to generate dynamic termination conditions for any algorithm that can be interpreted as an SI \ac{GRAND} decoder with a sub-optimal query order. The approach recovers the original \ac{GRAND} \ac{BLER} performance while preserving or enhancing the guesswork reduction. To demonstrate the effectiveness, we propose a new SI \ac{GRAND} variant, \ac{SyGRAND}, that, inspired by \ac{ORDEPT}, leverages syndrome information to find candidate codewords early. With the new termination condition, \ac{SyGRAND} achieves significant guesswork reduction compared to \ac{ORDEPT} without loss of \ac{BLER} performance in comparison to \ac{ORBGRAND}. To evaluate the method's effectiveness, we compare it to \ac{GCD}~\cite{maGuessingWhatNoise2024a}, which exploits codebook structure in a distinct way to reduce guesswork. Here we establish that if approximate orders are used, \ac{GRAND} algorithms can give identical or better \ac{BLER} performance to \ac{GCD} with significantly lower average guesswork.  

\section{Background}\label{sec:background}
\subsection{Channel}
We consider a transmission in which the data is protected by a binary linear \((n, k)\) code with \(m = n-k\) redundancy bits specified by its generator matrix \(G \in \{0,1\}^{k \times n}\) and parity check matrix \(H \in \{0, 1\}^{m \times n}\). Let 
\(
    \mathcal{C} = \{c^n \in \{0, 1\}^n : H c^n = 0^m\} \subset \{0, 1\}^n
\)
be the codebook of the code and \(r = k/n\) its rate.
Each message \(u^k\) is encoded to a codeword \(c^n = G u^k\) and sent over the communication channel.
At the receiver side, \acp{LLR} \(\ell^n \in \freal^n\) are input into the decoder.
The decoding algorithm works with any modulation as long as \acp{LLR} for the individual bits can be provided to the decoder. For the results in this paper, we assume that the codeword bits \(c_i\) for \(i \in [n] \coloneqq \{1, 2, \dots, n\}\) are transmitted over an \ac{BIAWGN} channel as \ac{BPSK} symbols \(x_i = (-1)^{c_i}\) resulting in \(y_i = x_i + z_i\) at the receiver side, where the noise samples \(z_i\) are independent normal distributed with noise variance \(\sigma^2 = (2 r \lEbNO)^{-1}\).
At the receiver, \acp{LLR} are calculated as \(\ell_i = 2y_i/\sigma^2\) and the hard-decision word \(y_{\text{hd}}^n\) is calculated as \(y_{\text{hd}, i} = \ind_{\{\ell_i < 0\}}\).

\subsection{GRAND}\label{sec:grand}
\ac{GRAND}~\cite{duffyCapacityAchievingGuessingRandom2019} iterates through potential noise patterns \(\tilde{z}^{n, (j)}\). For each \(\tilde{z}^{n, (j)}\), it subtracts the noise pattern from the hard decision received sequence \(y_{\text{hd}}^n\): 
\(y^{n, (j)} \coloneqq y_\text{hd}^n \ominus \tilde{z}^{n, (j)}\) and checks whether \(y^{n, (j)}\) is a valid codeword. If patterns are produced in decreasing order of likelihood, the first codeword found is an \ac{ML} codeword. To check codebook membership for linear codes, for each generated noise pattern \(\tilde{z}^{n, (j)}\), GRAND can calculate the syndrome of the received word with the noise pattern removed:
\begin{equation}\label{eqn:syndrome-calc}
    s^{m,(j)} = H y^{n, (j)} = H (y_\text{hd}^n \ominus \tilde{z}^{n, (j)}).
\end{equation}
If the syndrome is the all-zero vector, \(s^{m,(j)} = 0^m\), the guess \(\hat{c}^n = y^{n, (j)} = y_\text{hd}^n \ominus \tilde{z}^{n, (j)}\) is a valid codeword and returned as the decoding result. With each proposed decoding, \ac{SOGRAND} \cite{yuanSOGRAND25} can provide an accurate estimate of the probability that the codeword is correct or the correct decoding has not been found. Let
\begin{equation*}
    P_{Y^{n} | L^n}(y^{n,(j)} | \ell^n) = \prod_{i = 1}^n 
    \begin{cases}
        (1 + \exp(\ell_i))^{-1}, & y^{(j)}_i = 1 \\
        (1 + \exp(-\ell_i))^{-1}, & y^{(j)}_i = 0 \\
    \end{cases}
\end{equation*}
be the a posteriori probability of the \(j\)-th guessed word \(y^{n,(j)}\).
During guessing, the algorithm accumulates the probabilities of the guesses \(y^{n,(j)}\) as follows
\begin{equation}\label{eqn:calc-noise-prob}
    P_{\text{noise}} = \sum\limits_{j=1}^q P_{Y^{n} | L^n}(y^{n,(j)} | \ell^n), 
\end{equation}
where \(q\) denotes the number of guesses made.
Let \(\mathcal{L} \subset \mathcal{C}\) be a list of the codewords that \ac{GRAND} has found until the \(q\)-th guess. Following results in \cite{galligan2023upgrade,yuanSOGRAND25}, the probability that the correct codeword is not in the list is estimated as
\begin{equation}\label{eqn:soft-output}
    \hat{P}(C^n \notin \mathcal{L} | \ell^n) =
    \frac{
        (1 - P_{\text{noise}}) 2^{k-n}
    }{
        P_\mathcal{L}  + 
        (1 - P_{\text{noise}}) 2^{k-n}
    },
\end{equation}
where
\begin{equation}\label{eqn:calc-list-prob}
    P_\mathcal{L} = \sum_{w^n \in \mathcal{L}}
    P_{Y^{n} | L^n}(w^n | \ell^n),
\end{equation}
which only depends on the code dimensions and the likelihoods of queried sequences, both of which can be evaluated by the decoder.

\section{SyGRAND}\label{sec:sygrand}
We first discuss how codewords can be found early using the syndrome information, followed by the key innovation that terminates decoding early when the decoding result is sufficiently reliable. See Algorithm~\ref{alg:sygrand} for pseudo-code.

\begin{algorithm}[tpb]
\caption{SyGRAND($\theta, \mathcal{L}_\text{max}$) }
\label{alg:sygrand}
\KwIn{Parity-check matrix \(H\in\{0,1\}^{m\times n}\) and received LLR vector \(\ell^n\).}
\KwOut{Decoded codeword \(c^n_\textnormal{best}\)}
\BlankLine
For \(i \in [n]\), \(y_{\text{hd},i} \gets \ind_{\{\ell_i\leq 0\}}\)\;
\(P_\text{noise} \gets 0, \quad P_\mathcal{L} \gets 0, \quad \mathcal{L} \gets \{\}\)\;
\BlankLine
\ForEach{\(\tilde{z}^{n,(j)}\) in \(\textnormal{NoiseGenerator}(|\ell^n|)\)}{
    \(y^{n,(j)} \gets y_{\text{hd}}^n \oplus \tilde{z}^{n,(j)}\)\;
    \(P_\text{noise} \gets P_\text{noise} + P_{Y^{n}|L^n}(y^{n,(j)}|\ell^n)\)\;
    \(s^{m,(j)} \gets H\, y^{n,(j)}\)\;
    \If{\(s^{m,(j)} = 0^m\)}{
        \Return{\(y^{n,(j)}\)} \tcp*[r]{Codeword found}
    }
    \tcp{-----SyGRAND Extension-----}
    For \(r\in [n]\), \(g_r \gets \eqref{eqn:calc-g}\)\;
    \If{\(g_p = 1\) for a \(p \in [n]\)}{
        \(\hat{w}^n \gets y^{n,(j)} \oplus e_p^n\)\; 
        \If{\(\hat{w}^n \in \mathcal{L}\)}{
            \Continue
        }
        
        \tcp{New candidate codeword found}
        \(\mathcal{L} \gets \mathcal{L} \cup \{\hat{w}^n\}\)\;
        \(P_\mathcal{L} \gets P_\mathcal{L} + P_{Y^{n} | L^n}(\hat{w}^n | \ell^n)\)\;
        \(\hat{P}(C^n \notin \mathcal{L} | \ell^n) \gets \eqref{eqn:soft-output-sygrand}\)\;
        \If{\(\hat{P}(C^n \notin \mathcal{L} | \ell^n) \leq \theta\) or \(|\mathcal{L}| = \mathcal{L}_\textnormal{max}\)}{
            \Break
        }
    }
}
\Return{\(c^n_\textnormal{best} = \eqref{eqn:calc-cbest}\)}\;
\end{algorithm}

\subsection{Syndrome-Based Candidate Codewords Search}\label{sec:finding-candidates}

For each guess \(y^{j,(n)}\) of the \ac{GRAND} decoder, SyGRAND checks whether the syndrome \(s^{m,(j)}\) matches one of the columns of \(H\). The check can be easily realized with an additional boolean logic:
\begin{equation}\label{eqn:calc-g}
    g_r = \bigwedge_{i=1}^m (H_{r,i} = s^{m,(j)}_i)
\end{equation}
for all \(r \in [n]\).
The calculation of \(g^n\) can be understood as the matrix multiplication \((s^{m,(j)})^T H\), where the binary multiplication and addition are replaced by the binary equality and binary multiplication operation, respectively.
If \(g_p = 1\), for one \(p \in [n]\), a valid codeword can be found by applying an additional bit flip at position \(p\) to the current guess~\cite{hadavianOrderedReliabilityDirect2023},
\begin{equation}\label{eqn:bit-flip}
    \hat{w}^n = y^{j,(n)} \oplus e_p^n,
\end{equation}
where \(e_p^n\) is the \(p\)-th unit vector, which is \(1\) at position \(p\) and \(0\) otherwise.
Eq. \eqref{eqn:bit-flip} follows from
\[
    H \hat{w}^n 
    = H y^{j,(n)} \oplus H e_p^n
    = s^{m,(j)} \oplus H_{:, p}
    = 2 \cdot s^{m,(j)} = 0^m,
\]
using the linearity of the matrix multiplication.

Eq. \eqref{eqn:calc-g} and \eqref{eqn:bit-flip} effectively carry out syndrome decoding~\cite[Sec.~3.1.4]{ryan2009channel} during every guess, 
which can be interpreted as bounded distance decoding~\cite[Sec.~1.8.1]{moonErrorCorrectionCoding2005} with Hamming spheres
\(
    \mathcal{S}_1(\hat{c}^n) = \{v^n : d_\text{H}(\hat{c}^n, v^n) \leq 1\}
\)
around every codeword  \(\hat{c}^n\). 
\begin{figure}
    \centering
    \includegraphics[width=0.9\columnwidth]{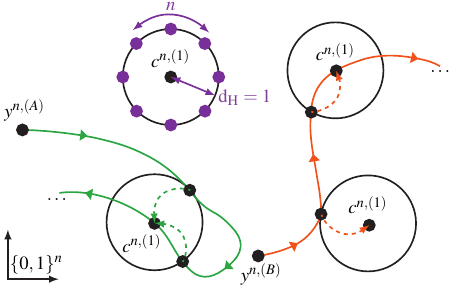}
    \caption{Binary words are visualized in 2D, where the distance captures the Hamming Distance between words. To decode the hard decision received words \(y^{n, (A)}\) and \(y^{n, (B)}\) \ac{GRAND} subtracts a sequence of noise patterns informed by the reliabilities (green and orange path). Decoding terminates once the path reaches a codeword. \ac{SyGRAND} leverages the syndrome that \ac{GRAND} calculates at each guess, allowing the decoder to identify candidate codewords if it reaches any of the \(n+1\) words of its Hamming sphere of radius 1 (purple sphere, upper left), which reduces guesswork (decoding of \(y^{n, (A)}\)). However, the decoding may not be the \ac{ML} codeword (decoding of \(y^{n, (B)}\)), leading to a degradation in \ac{BLER}. To ameliorate performance loss, dynamic list decoding terminates once the \ac{SOGRAND} estimated list error probability is below a threshold.}
    \label{fig:visualization-sygrand}
\end{figure}
Fig.~\ref{fig:visualization-sygrand} visualizes the codeword space.
\ac{SyGRAND} finds a candidate codeword if the underlying \ac{GRAND} decoder guesses a word in the union of the Hamming spheres \(\cup_{\hat{c}^n \in \mathcal{C}} \mathcal{S}_1(\hat{c}^n)\) forming a nonlinear code of rate
\[
    r 
    = \frac{\log_2(2^k \cdot |\mathcal{S}_1(\hat{c}^n)|)}{n}
    = \frac{k + \log_2(n+1)}{n}.
\]
In the low \ac{SNR} regime with substantial noise, one can expect that the average guesswork that \ac{GRAND} needs to find the first candidate codeword is reduced by a factor of \(\log_2(n+1)\). 

\subsection{List Decoding and Decoding Termination}\label{sec:filtering-candidates}
While the syndrome-based candidate codeword search finds codewords quickly, it does so by changing \ac{GRAND}'s guessing order. An optimal \ac{GRAND} algorithm iterates through the noise patterns in approximately decreasing order of probability, ensuring that the first codeword found is the most likely. If codewords are identified early via an additional bit flip, this property is no longer fulfilled, and the \ac{ML}-codeword might appear later in the guessing order, resulting in a \ac{BLER} performance loss (see Fig.~\ref{fig:visualization-sygrand}). One way to mitigate the performance loss is \ac{GRAND} list decoding~\cite{abbasListGRANDPracticalWay2023}, which continues guesswork after the first codeword is found and collects all valid codewords in a list \(\mathcal{L}\). After the guesswork is stopped by a termination condition, the most likely codeword in the list
\begin{equation}\label{eqn:calc-cbest}
    c^n_\text{best} =
    \operatorname{argmax}_{w^n \in \mathcal{L}} P_{Y^{n}|L^n}(w^n|\ell^n)
\end{equation}
is returned as the decoding result.

While list decoding eliminates the performance loss, it counteracts the guesswork reduction of the algorithm. 
The solution considered here is a termination condition based on \ac{SOGRAND}, which estimates the likelihood that the correct codeword has been found after each addition to the list. If the likelihood exceeds a predefined threshold, the decoding terminates. Since \ac{SOGRAND} is highly accurate~\cite{yuanSOGRAND25}, unnecessary guesswork can be efficiently avoided. This results in a decoder with a dynamic list size, adapting the guesswork to each specific decoding. In particular, for each candidate codeword found through an additional bit flip, \ac{SyGRAND} first checks whether it is already part of the codeword list \(\mathcal{L} \subset \mathcal{C}\). If not, the candidate is added to the list%
\footnote{As candidate codewords are generated through an additional bit flip, they may be encountered multiple times, making it necessary to prevent double counting.}
\(\mathcal{L}\) and the likelihood that the correct codeword \(c^n\) is not in the list is calculated as
\begin{equation}\label{eqn:soft-output-sygrand}
    \hat{P}(c^n \notin \mathcal{L} | \ell^n) =
    \frac{
        (1 - (P_{\text{noise}} + P_\mathcal{L})) 2^{k-n}
    }{
        P_\mathcal{L}  + 
        (1 - (P_{\text{noise}} + P_\mathcal{L})) 2^{k-n}
    },
\end{equation}
where \(P_{\text{noise}}\) and \(P_\mathcal{L}\) are calculated using Eq. \eqref{eqn:calc-noise-prob} and \eqref{eqn:calc-list-prob}, respectively.
If the termination condition
\[
    \hat{P}(c^n \notin \mathcal{L} | \ell^n) \leq \theta \qquad \text{or} \qquad |\mathcal{L}| = \mathcal{L}_\text{max},
\]
is fulfilled, decoding stops, and the final decoding result is determined using Eq. \eqref{eqn:calc-cbest}.
The maximal error probability \(\theta \in [0, 1]\) and list size \(\mathcal{L}_\text{max} \in \mathbb{N}\) are parameters.
Note that in Eq. \eqref{eqn:soft-output-sygrand}, in contrast to \ac{SOGRAND} in Eq. \eqref{eqn:soft-output}, the noise probability \(P_{\text{noise}}\) is increased by \(P_\mathcal{L}\) as the candidate codewords in \(\mathcal{L}\) do not correspond to guessed noise patterns.
In addition to candidate codewords, \ac{SyGRAND} identifies a valid codeword if the syndrome \(s^{m,(j)}\) in Eq. \eqref{eqn:syndrome-calc} is all-zero, in which case the codeword is output without continuing list decoding.

\subsection{Decoder Parameters}\label{sec:parameter-optimization}
In the following, we refer to the \ac{SyGRAND} decoder with parameters \(\theta\) and a \(\mathcal{L}_\text{max}\) as \ac{SyGRAND}(\(\theta, \mathcal{L}_\text{max}\)).
\(\theta\) and  \(\mathcal{L}_\text{max}\) are per-code hyperparameters: a higher \(\theta\) and smaller \(\mathcal{L}_\text{max}\) result in earlier termination, thereby reducing average guesswork but potentially degrading \ac{BLER} performance.
Let \(\text{BLER}_\text{Ref}(\lEbNO)\) be the \ac{BLER} curve of a reference decoder and \(\text{BLER}_\text{SyGRAND(\(\theta, \mathcal{L}_\text{max}\))}(\lEbNO)\) the \ac{BLER} curve of \ac{SyGRAND}(\(\theta, \mathcal{L}_\text{max}\)) for a code \(\mathcal{C}\):
The parameters are optimized through two sequential one-dimensional optimizations:
\begin{enumerate}[leftmargin=*]
    \item Set \(\theta = 0\), i.e., no decoding termination via \(\hat{P}(c^n \notin \mathcal{L} | \ell^n)\) and find
    \begin{multline*}
        \mathcal{L}_\text{max}^\star \coloneqq \min \Big\{ 
            \mathcal{L}_\text{max} \in \mathbb{N}: \forall \lEbNO:\\
            \text{BLER}_\text{SyGRAND(\(0, \mathcal{L}_\text{max}\))}(\lEbNO)
            \leq
            \text{BLER}_\text{Ref}(\lEbNO)
            \Big\}.
    \end{multline*}
    \item Set \(\mathcal{L}_\text{max} = \mathcal{L}_\text{max}^\star\) and find
    \begin{multline*}
        \theta^\star \coloneqq \max \Big\{
            \theta \in [0, 1]: \forall \lEbNO:\\
            \text{BLER}_\text{SyGRAND(\(\theta, \mathcal{L}_\text{max}^\star\))}(\lEbNO)
            \leq
            \text{BLER}_\text{Ref}(\lEbNO)
            \Big\}.
    \end{multline*}
\end{enumerate}

\section{Performance Evaluation}
\begin{figure}
    \centering
    \includegraphics[width=0.95\columnwidth]{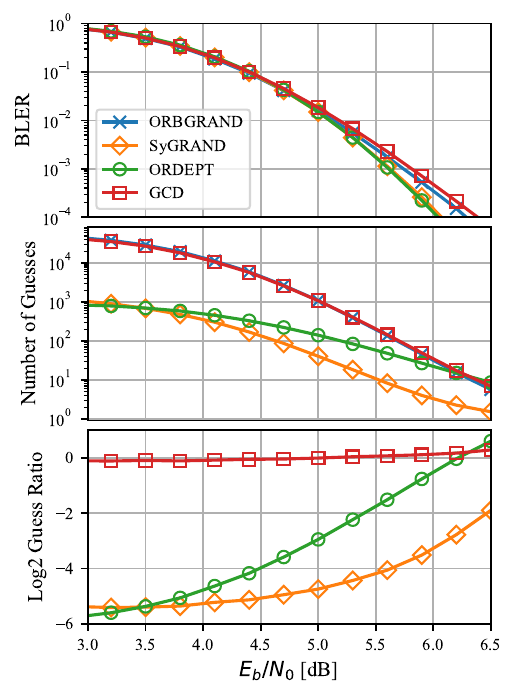}
    \caption{eBCH (256, 239). Top panel: BLER vs \(\lEbNO\). Middle panel: average guesswork vs. \(\lEbNO\). Bottom panel: average guesswork divided by average guesswork of the \ac{ORBGRAND}, plotted in \(\log_2\)-scale. 
    Parameters: \ac{SyGRAND}: \(\theta= 0.7, \mathcal{L}_\text{max} = 5\), \ac{ORDEPT}: \(t=450, c_\text{max}=5\).}
    \label{fig:ebch256-239}
\end{figure}

\begin{figure}
    \centering
    \includegraphics[width=0.95\columnwidth]{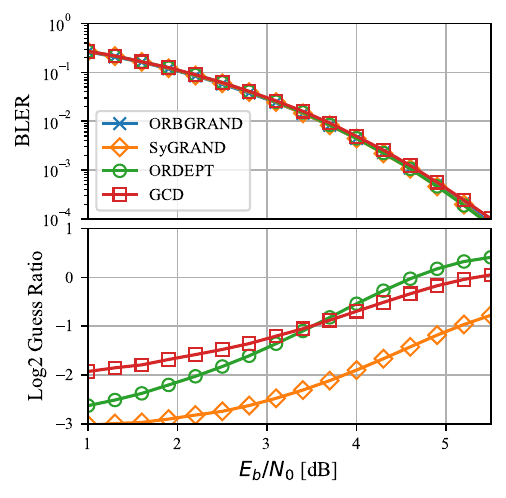}
    \caption{eBCH (32, 21). Parameters: \ac{SyGRAND}: \(\theta= 0.71, \mathcal{L}_\text{max} = 3\), ORDEPT: \(t=50, c_\text{max}=3\).}
    \label{fig:ebch-32-21}
\end{figure}

\begin{figure}
    \centering
    \includegraphics[width=0.95\columnwidth]{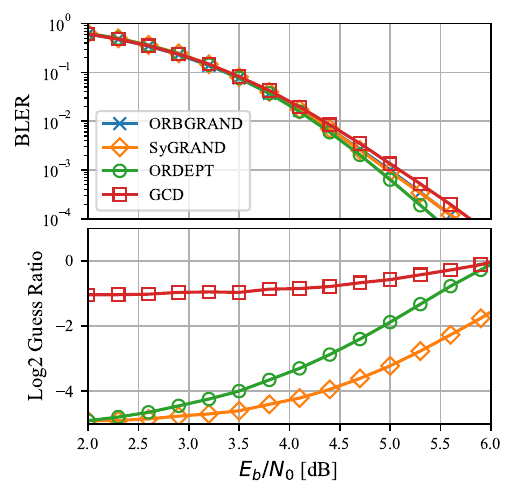}
    \caption{CA-Polar (128, 110). Parameters: \ac{SyGRAND}: \(\theta= 0.54, \mathcal{L}_\text{max} = 3\), ORDEPT: \(t=3500, c_\text{max}=3\).}
    \label{fig:ebch-128-110}
\end{figure}

We compare \ac{SyGRAND} with \ac{ORBGRAND}, \ac{ORDEPT} and \ac{GCD} in terms of \ac{BLER} performance and guesswork reduction. All decoders use 1-line \ac{ORBGRAND}\cite{duffyOrderedReliabilityBits2022} for their generator. 
For even-weight codes, \ac{ORBGRAND}, \ac{SyGRAND} and \ac{ORDEPT} only query codeword guesses \(y^{n,(j)}\) with even parity. This reduction is not available to \ac{GCD}.
\ac{ORDEPT}'s parameters \(t\) and \(c_\text{max}\) are optimized using the same procedure as \ac{SyGRAND} described in Sec.~\ref{sec:parameter-optimization}.
\ac{SyGRAND}'s and \ac{ORDEPT}'s parameter optimization both use 1-line \ac{ORBGRAND}'s \ac{BLER} performance as their reference curve.
We assess the \ac{eBCH} (256, 239) code in Fig.~\ref{fig:ebch256-239}, the \ac{eBCH} (32, 21) in Fig.~\ref{fig:ebch-32-21}, and the \ac{CAPolar} (128, 110) code, with the \(11\)-bit CRC from 5G new radio, in Fig.~\ref{fig:ebch-128-110}.

The upper plot of each figure shows \ac{BLER} vs \(\lEbNO\), demonstrating that, by design, \ac{SyGRAND} performs at least as well as \ac{ORBGRAND}.
The middle plot in Fig.~\ref{fig:ebch256-239} shows average guesswork in \(\log_{10}\)-scale. 
To clarify the comparison across decoders, the average guesswork of each decoder is divided by that of \ac{ORBGRAND} and shown in \(\log_2\)-scale as the lower plot in all figures. A negative \(\log_2\)-ratio means that the respective decoder reduces guesswork compared to \ac{ORBGRAND}. One can see that \ac{SyGRAND} outperforms \ac{ORDEPT} in terms of guesswork reduction, and the \ac{ORDEPT} advantage vanishes at higher \(\lEbNO\).

\ac{SyGRAND} consistently outperforms \ac{GCD} in terms of guesswork for all codes across the entire \(\lEbNO\) range, requiring up to a factor \(32\) fewer average guesses. This indicates that leveraging code-book information to inform a query order can be more effective in a noise-centric approach than a codebook based one.

\section{Conclusion}
We introduce a technique to eliminate the \ac{BLER} performance loss for \ac{GRAND} decoders with suboptimal guessing order through dynamically sized list decoding. List decoding terminates once a sufficiently reliable decoding has been identified, minimizing additional guesswork overhead. Building on that idea, we developed \ac{SyGRAND}, which leverages syndrome information to find candidate codewords early in the guessing order. Results indicate that \ac{SyGRAND} significantly reduces guesswork across the entire \(\lEbNO\) range without any \ac{BLER} performance loss and that \ac{SyGRAND} significantly outperforms \ac{GCD} in terms of average guesswork reduction.
Beyond \ac{SyGRAND}, the framework applies to any decoder that can be understood through a \ac{SOGRAND} lens, opening the door for future guesswork reduction techniques where a suboptimal guessing order is no longer a limiting factor.

\section{Acknowledgment}
This material is based upon work supported by the Defense Advanced Research Projects Agency (DARPA) under Contract Numbers HR00112120008 and HR001124C0403.

\bibliography{main.bbl}

\begin{thebibliography}{10}
\providecommand{\url}[1]{#1}
\csname url@samestyle\endcsname
\providecommand{\newblock}{\relax}
\providecommand{\bibinfo}[2]{#2}
\providecommand{\BIBentrySTDinterwordspacing}{\spaceskip=0pt\relax}
\providecommand{\BIBentryALTinterwordstretchfactor}{4}
\providecommand{\BIBentryALTinterwordspacing}{\spaceskip=\fontdimen2\font plus
\BIBentryALTinterwordstretchfactor\fontdimen3\font minus
  \fontdimen4\font\relax}
\providecommand{\BIBforeignlanguage}[2]{{%
\expandafter\ifx\csname l@#1\endcsname\relax
\typeout{** WARNING: IEEEtran.bst: No hyphenation pattern has been}%
\typeout{** loaded for the language `#1'. Using the pattern for}%
\typeout{** the default language instead.}%
\else
\language=\csname l@#1\endcsname
\fi
#2}}
\providecommand{\BIBdecl}{\relax}
\BIBdecl

\bibitem{duffyCapacityAchievingGuessingRandom2019}
K.~R. Duffy, J.~Li, and M.~M{\'e}dard, ``Capacity-achieving guessing random
  additive noise decoding,'' \emph{IEEE Trans. Inf. Theory}, vol.~65, no.~7,
  pp. 4023--4040, Jul. 2019.

\bibitem{cohenAESErrorCorrection2023}
A.~Cohen, R.~G.~L. D'Oliveira, K.~R. Duffy, J.~Woo, and M.~Medard, ``{{AES}} as
  {{Error Correction}}: {{Cryptosystems}} for {{Reliable Communication}},''
  \emph{IEEE Commun. Lett.}, vol.~27, no.~8, pp. 1964--1968, 2023.

\bibitem{An22GRANDMO}
W.~An, M.~M\'edard, and K.~R. Duffy, ``Keep the bursts and ditch the
  interleavers,'' \emph{IEEE Trans. Commun.}, vol.~70, no.~6, pp. 3655--3667,
  2022.

\bibitem{abbas2020grand}
S.~M. Abbas, T.~Tonnellier, F.~Ercan, and W.~J. Gross, ``{High-Throughput VLSI
  Architecture for GRAND},'' in \emph{IEEE Workshop on Sig. Proc. Sys.}, 2020,
  pp. 681--693.

\bibitem{abbas2021grandmo}
S.~M. Abbas, M.~Jalaleddine, and W.~J. Gross, ``{High-Throughput VLSI
  Architecture for GRAND Markov Order},'' in \emph{IEEE Workshop Sig. Proc.
  Sys.}, 2021.

\bibitem{Riaz21}
A.~Riaz, V.~Bansal, A.~Solomon, W.~An, Q.~Liu, K.~Galligan, K.~R. Duffy,
  M.~M\'edard, and R.~T. Yazicigil, ``Multi-code multi-rate universal maximum
  likelihood decoder using {GRAND},'' in \emph{IEEE ESSCIRC}, 2021, pp.
  239--246.

\bibitem{Burg24}
L.~D. Blanc, V.~Herrmann, Y.~Ren, C.~Müller, A.~T. Kristensen, A.~Levisse,
  Y.~Shen, and A.~Burg, ``A {GRANDAB} decoder with 8.48 {Gbps} worst-case
  throughput in 65nm {CMOS},'' in \emph{IEEE ESSERC}, 2024, pp. 685--688.

\bibitem{solomonSoftMaximumLikelihood2020b}
A.~Solomon, K.~R. Duffy, and M.~M{\'e}dard, ``Soft maximum likelihood decoding
  using {{GRAND}},'' in \emph{{{IEEE ICC}}}, Jun. 2020, pp. 1--6.

\bibitem{duffyOrderedReliabilityBits2022}
K.~R. Duffy, W.~An, and M.~M{\'e}dard, ``Ordered reliability bits guessing
  random additive noise decoding,'' \emph{IEEE Trans. Signal Process.},
  vol.~70, pp. 4528--4542, 2022.

\bibitem{an2023soft}
W.~An, M.~M{\'e}dard, and K.~R. Duffy, ``Soft decoding without soft demapping
  with {ORBGRAND},'' in \emph{IEEE ISIT}, 2023, pp. 1080--1084.

\bibitem{duffyUsingChannelCorrelation2023a}
K.~R. Duffy, M.~Grundei, and M.~M{\'e}dard, ``Using channel correlation to
  improve decoding - {{ORBGRAND-AI}},'' in \emph{{{IEEE GLOBECOM}}}, Dec. 2023,
  pp. 3585--3590.

\bibitem{condo2021_highperformance}
C.~Condo, V.~Bioglio, and I.~Land, ``High-performance low-complexity error
  pattern generation for {ORBGRAND} decoding,'' in \emph{{IEEE GLOBECOM}},
  2021.

\bibitem{condo2021fixed}
C.~Condo, ``A fixed latency {ORBGRAND} decoder architecture with {LUT}-aided
  error-pattern scheduling,'' \emph{IEEE Trans. Circuits Sys. I: Regular
  Papers}, vol.~69, no.~5, pp. 2203--2211, 2022.

\bibitem{abbas2022orbgrand}
S.~M. Abbas, T.~Tonnellier, F.~Ercan, M.~Jalaleddine, and W.~J. Gross,
  ``{High-Throughput and Energy-Efficient VLSI Architecture for Ordered
  Reliability Bits GRAND},'' \emph{IEEE Trans. on VLSI Sys.}, vol.~30, no.~6,
  2022.

\bibitem{Riaz23}
A.~Riaz, A.~Yasar, F.~Ercan, W.~An, J.~Ngo, K.~Galligan, M.~M\'edard, K.~R.
  Duffy, and R.~T. Yazicigil, ``A sub-0.8p{J}/b 16.3{G}bps/mm$^2$ universal
  soft-detection decoder using {ORBGRAND} in 40nm {CMOS},'' in \emph{IEEE
  ISSCC}, 2023.

\bibitem{Riaz25}
A.~Riaz, A.~Yasar, F.~Ercan, W.~An, J.~Ngo, K.~Galligan, M.~Médard, K.~R.
  Duffy, and R.~T. Yazicigil, ``A sub-0.8-{pJ}/bit universal soft-detection
  decoder using {ORBGRAND},'' \emph{IEEE J. Solid-State Circuits}, vol. to
  appear, 2025.

\bibitem{abbasListGRANDPracticalWay2023}
S.~M. Abbas, M.~Jalaleddine, and W.~J. Gross, ``List-{{GRAND}}: {{A}} practical
  way to achieve maximum likelihood decoding,'' \emph{IEEE Trans. Very Large
  Scale Integr. (VLSI) Syst.}, vol.~31, pp. 43--54, Jan. 2023.

\bibitem{galligan2023upgrade}
K.~Galligan, P.~Yuan, M.~M\'edard, and K.~R. Duffy, ``Upgrade error detection
  to prediction with {GRAND},'' in \emph{IEEE GLOBECOM}, 2023.

\bibitem{yuanSOGRAND25}
P.~Yuan, M.~Médard, K.~Galligan, and K.~R. Duffy, ``Soft-output {(SO) GRAND}
  and iterative decoding to outperform {LDPC} codes,'' \emph{IEEE Trans.
  Wireless Commun.}, vol. to appear, 2025.

\bibitem{rowshanConstrainedErrorPattern2022a}
M.~Rowshan and J.~Yuan, ``Constrained error pattern generation for {{GRAND}},''
  in \emph{{{IEEE ISIT}}}, Jun. 2022, pp. 1767--1772.

\bibitem{rowshanLowComplexityGRANDSegmentation2023}
------, ``Low-complexity {{GRAND}} by segmentation,'' in \emph{{{IEEE
  GLOBECOM}}}, Dec. 2023, pp. 6145--6151.

\bibitem{rowshanSegmentedGRANDComplexity2025a}
------, ``Segmented {{GRAND}}: {{Complexity}} reduction through sub-pattern
  combination,'' \emph{IEEE Trans. Commun.}, vol. to appear, 2025.

\bibitem{isitpaper}
L.~Rapp, J.~Feng, M.~M{\'e}dard, and K.~R. Duffy, ``A balanced tree
  transformation to reduce {GRAND} queries,'' \emph{arXiv:2503.19033}, 2025.

\bibitem{hadavianOrderedReliabilityDirect2023}
R.~Hadavian, D.~Truhachev, K.~{El-Sankary}, H.~Ebrahimzad, and H.~Najafi,
  ``Ordered reliability direct error pattern testing ({{ORDEPT}}) algorithm,''
  in \emph{{{IEEE GLOBECOM}}}, 2023, pp. 6983--6988.

\bibitem{maGuessingWhatNoise2024a}
X.~Ma, ``Guessing what, noise or codeword?'' in \emph{{{IEEE ITW}}}, Nov. 2024,
  pp. 460--465.

\bibitem{ryan2009channel}
W.~E. Ryan and S.~Lin, \emph{Channel {{Codes}}: {{Classical}} and
  {{Modern}}}.\hskip 1em plus 0.5em minus 0.4em\relax Cambridge University
  Press, 2009.

\bibitem{moonErrorCorrectionCoding2005}
T.~K. Moon, \emph{Error {{Correction Coding}}: {{Mathematical Methods}} and
  {{Algorithms}}}.\hskip 1em plus 0.5em minus 0.4em\relax Hoboken, NJ:
  Wiley-Interscience, 2005.

\end{thebibliography}

\end{document}